\documentclass[preprint,showpacs,preprintnumbers,amsmath,amssymb]{revtex4}

\usepackage{graphicx}
\usepackage{dcolumn}
\usepackage{bm}

\begin{document}

\preprint{APS/123-QED}

\title{Elliptical flow and isospin effects in heavy-ion collisions 
at intermediate energies\\}

\author{Sanjeev Kumar}
\author{Suneel Kumar}%
 \email{suneel.kumar@thapar.edu}
\affiliation{%
School of Physics and Material Science, Thapar University, Patiala-147004, Punjab (India)\\
}%

\author{Rajeev K. Puri}
\affiliation{
Department of Physics, Panjab University, Chandigarh-160014 (India)\\
}%

\date{\today}

\begin{abstract}
The elliptical flow of fragments is studied for different systems at incident 
energies between 50 and 1000 MeV/nucleon using Isospin-dependent Quantum 
Molecular Dynamics (IQMD) Model. Our findings reveal that 
elliptical flow shows a 
transition from positive (in-plane) to negative (out-of-plane) value in the 
mid-rapidity region at certain incident energy, 
known as transition energy. This transition energy is found to depend on the 
model ingredients, size of the fragments, 
composite mass of the reacting system as well as on the impact parameter 
of the reaction. A reasonable agreement is observed for the excitation 
function of elliptical flow between the data and our calculations. 
Interestingly, the transition energy is found to exhibit a power law 
mass dependence.\\ 
\end{abstract}

\maketitle

\section{Introduction}
The information about the nature of equation of state is still one of 
the burning topic of present day nuclear physics research in general and heavy-ion
collisions in particular. A quite good progress has been made in the recent years 
in determining the nuclear equation of state from heavy-ion reactions 
\cite{Dani02,Aich91}. Among different observables, collective flow enjoys a special status. 
This is due to its sensitive response to the model ingredients that define equation of state.
A lot of theoretical and experimental efforts have been made in studying the 
collective flow in heavy-ion collisions \cite{West93,Luka05,Andr05,Chen06, Luka04,
Kuma98,Sood06}.
This collective motion of the particles in heavy-ion collision can be studied via
directed and elliptical flows. The directed flow, which measures the 
collective motion of the particles in the reaction plane,
has been studied extensively at BEVALAC, SIS and AGS energies \cite{Gyul83}. 
This flow is reported to diminish at higher incident energies
due to the large beam rapidity. Therefore, elliptical flow \cite{Sorg97} is much more
suited at these incident energies.
The elliptical flow describes the eccentricity of an ellipse like distribution. 
Quantitatively, it is the difference between the major and minor axis. 
The orientation of the major axis is confined to azimuthal angle 
$\phi$ or $\phi$+$\frac{\pi}{2}$ for 
ellipse like distribution. The major axis lies within the reaction plane
for $\phi$; while $\phi$+$\frac{\pi}{2}$ 
indicates that the orientation of the ellipse is perpendicular to the reaction plane, which is the case for 
squeeze out flow and may be expected at mid rapidity \cite{Volo96}.
Therefore, the elliptical flow is defined by the
second order Fourier coefficient
from the azimuthal distribution of detected particles at mid rapidity. Mathematically,
\begin{equation}
\frac{dN}{d\phi} = p_0(1 + 2v_1Cos\phi + 2v_2Cos2\phi).
\end{equation}
Here $\phi$ is the azimuthal angle between the transverse momentum of the particle and 
reaction plane.
The positive value of
the elliptical flow $<Cos2\phi>$ reflects an in-plane emission, whereas, out-of
plane emission is reflected by its negative value.
The reason for the anisotropic flow is orthogonal asymmetry in the configuration space (non-central collisions) and
re-scattering. In the case of elliptical flow, the initial "ellipticity" of the overlap 
zone is usually characterized
by a quantity $\epsilon = \frac{(<y^2-x^2>)}{(<y^2+x^2>)}$, assuming the reaction 
plane being xz. As the system 
expands, spatial anisotropy decreases. From the above discussion, it is clear that 
the second order flow (elliptical flow) is 
better candidate for determining the nuclear equation of state compared to 
first order sideward flow (directed flow). \\
In recent years, several experimental groups have measure
the elliptical flow. The FOPI, INDRA and PLASTIC BALL collaborations \cite{Luka05,Andr05} are 
actively involved in 
measuring the excitation function of elliptical flow from Fermi energies to relativistic energies.
In most of these studies, $_{79}Au^{197}~+~_{79}Au^{197}$ reaction 
has been taken \cite{Luka05,Andr05}.
Interestingly, a change in the elliptical flow was reported from 
positive to negative values around
100 MeV/nucleon. Both the mean field and two-body binary collisions play 
an important role in this energy domain. 
The mean field is supposed to play a dominant role at low incident energies. The binary
collisions starts dominating the physics gradually. 
A detailed study of the excitation function of elliptical flow in entire energy region
can provide a useful information about the nucleon-nucleon interactions related to the 
nuclear equation of state.\\
As discussed above, lots of attempts have already been made in the literature to 
explore different 
aspects of directed
sideward flow. 
In this paper, we attempt to study the different aspects of elliptical flow $v_2$.\\
For the present study, Isospin-dependent Quantum Molecular Dynamics (IQMD) model is used 
to generate the phase space of nucleons. 
The article is organized as follow:
we discuss the model briefly in section-II. The results are discussed in section-III and we summarize
the results in section-IV. \\
\section{ISOSPIN-dependent QUANTUM MOLECULAR DYNAMICS (IQMD) MODEL}
The isospin-dependent quantum molecular dynamics (IQMD)\cite{Hart89} model treats different charge states of
nucleons, deltas and pions
explicitly\cite{Hart89}, as inherited from the VUU model \cite{Krus85}. 
The IQMD model has been used successfully
for the analysis of large number of observables from low to relativistic energies. 
The isospin degree of
freedom enters into the calculations via symmetry potential, cross-sections and
Coulomb interaction\cite{Krus85}.
The details about the elastic and inelastic cross-sections
for proton-proton and neutron-neutron collisions can be found in Ref.\cite{Hart89}. \\
In this model,  baryons are represented by Gaussian-shaped density distributions
\begin{equation}
f_i(\vec{r},\vec{p},t) = \frac{1}{\pi^2\hbar^2}\cdot e^{-(\vec{r}-\vec{r_i}(t))^{2}\frac{1}{2L}}\cdot e^{-(\vec{p}-\vec{p_i}(t))^{2}\frac{2L}{\hbar^2}}.
\end{equation}
Nucleons are initialized in a sphere with radius $R= 1.12 A^{1/3}$ fm, in accordance with the liquid drop model. Each nucleon occupies a volume of $h^3$, so that phase space is uniformly filled. The initial momenta are randomly chosen between 0
and Fermi momentum($p_F$). The nucleons of target and projectile
interact via two and three-body Skyrme forces, Yukawa potential, Coloumb interactions and momentum-dependent interactions. 
In addition to the use of explicit charge states of all baryons and mesons a symmetry potential between 
protons and neutrons corresponding to the Bethe- Weizsacker mass formula has been included.\\
The hadrons propagate using Hamilton equations of motion:
\begin{equation}
\frac{d\vec{r_i}}{dt}~=~\frac{d\it{\langle~H~\rangle}}{d\vec{p_i}}~~;~~\frac{d\vec{p_i}}{dt}~=~-\frac{d\it{\langle~H~\rangle}}{d\vec{r_i}},
\end{equation}
with
\begin{eqnarray}
\langle~H~\rangle&=&\langle~T~\rangle+\langle~V~\rangle\nonumber\\
&=&\sum_{i}\frac{p_i^2}{2m_i}+
\sum_i \sum_{j > i}\int f_{i}(\vec{r},\vec{p},t)V^{\it ij}({\vec{r}^\prime,\vec{r}})\nonumber\\
& &\times f_j(\vec{r}^\prime,\vec{p}^\prime,t)d\vec{r}d\vec{r}^\prime d\vec{p}d\vec{p}^\prime .
\end{eqnarray}
 The baryon-baryon potential $V^{ij}$, in the above relation, reads as:
\begin{eqnarray}
V^{ij}(\vec{r}^\prime -\vec{r})&=&V^{ij}_{Skyrme}+V^{ij}_{Yukawa}+V^{ij}_{Coul}+V^{ij}_{mdi}+V^{ij}_{sym}\nonumber\\
&=&\left(t_{1}\delta(\vec{r}^\prime -\vec{r})+t_{2}\delta(\vec{r}^\prime -\vec{r})\rho^{\gamma-1}
\left(\frac{\vec{r}^\prime +\vec{r}}{2}\right)\right)\nonumber\\
& & +~t_{3}\frac{exp(|\vec{r}^\prime-\vec{r}|/\mu)}{(|\vec{r}^\prime-\vec{r}|/\mu)}~+~\frac{Z_{i}Z_{j}e^{2}}{|\vec{r}^\prime -\vec{r}|}\nonumber\\
& & +t_{4}\ln^2[t_{5}(\vec{{p}_{i}}^\prime-\vec{p})^{2}+1]\delta(\vec{r}^\prime -\vec{r})\nonumber\\
& &+t_{6}\frac{1}{\varrho_0}T_3^{i}T_3^{j}\delta(\vec{r_i}^\prime -\vec{r_j}).
\label{s1}
\end{eqnarray}
Here $Z_i$ and $Z_j$ denote the charges of $i^{th}$ and $j^{th}$ baryon, and $T_3^i$, $T_3^j$ are their respective $T_3$
components (i.e. 1/2 for protons and -1/2 for neutrons). Meson potential consists of Coulomb interaction only.
The parameters $\mu$ and $t_1,.....,t_6$ are adjusted to the real part of the nucleonic optical potential. For the density
dependence of nucleon optical potential, standard Skyrme-type parameterization is employed.
The momentum dependence $V^{ij}_{mdi}$ of the N-N interactions, which may optionally be used in IQMD, is fitted to
experimental data in the real part of nucleon optical potential.
The choice of equation of state (or compressibility) is still controversial one. Many studies
advocate softer matter, whereas, much more believe the matter to be harder in
nature \cite{Krus85,Mage00}. As noted 
\cite{Zhan02}, elliptical flow is unaffected by the choice of equation of state.
For the present analysis, a hard (H) and hard momentum dependent (HMD) equation of state, 
has been employed alongwith standard energy dependent cross-section. \\
\section{Results and Discussion}
We, here, perform a complete systematic study for the mass range between 80 and 394 
units and over 
full range of the impact parameter. 
We here simulate the reactions of $_{20}Ca^{40}~+~ _{20}Ca^{40},~ 
_{28}Ni^{58}~+~ _{28}Ni^{58},~_{41}Nb^{93}~+~ _{41}Nb^{93},
~ _{54}Xe^{131}~+
~_{54}Xe^{131}$ and $_{79}Au^{197}~+~_{79}Au^{197}$ at incident energies 
between 50 and 1000 MeV/nucleon. In addition, the reactions of 
$_{40}Zr^{96}~+~_{40}Zr^{96}$ and  
$_{44}Ru^{96}~+~_{44}Ru^{96}$ are also simulated to check the isospin effects 
explicitly. 
As noted in Ref.\cite{Lehm93}, the relativistic effects do not play
role at these incident energies and the intensity of sub-threshold particle production is 
very small. The phase space generated by the IQMD model has been analyzed using 
the minimum spanning
tree (MST) \cite{Aich91,Verm09} method. The MST method binds two nucleons in a fragment if their distance is less than 4 fm. In
recent years, several improvements have also been suggested \cite{Sing00a}. One 
of the improvements is to also imply momentum cut of the order of Fermi 
momentum. This method is dubbed as MSTM method. The entire calculations are performed 
at $t = 200$ fm/c. This time is chosen by keeping in view the saturation of 
the collective flow \cite{Sood06}.\\
The elliptical flow is defined as the average difference between 
the square of the x and y components of the particle's
transverse momentum. Mathematically, it can be written as:
\begin{equation}
v_2 = \langle\frac{p_x^2 - p_y^2}{p_x^2 + p_y^2}\rangle,
\end{equation}
where $p_x$ and $p_y$ are the x and y components of the momentum. The 
$p_x$ is in the reaction plane, while, $p_y$ is 
perpendicular to the reaction plane.\\
A positive value of the elliptical flow describes the eccentricity of an ellipse like 
distribution and indicates
in-plane enhancement of the particle emission i.e. a rotational behavior. 
On the other hand, a negative value of $v_2$ shows the squeeze out effects 
perpendicular to the reaction plane. Obviously,
zero value corresponds to an
isotropic distribution in the
transverse plane. The $v_2$ is generally extracted from the mid rapidity region. 
The particles corresponding to
($Y_{c.m.}/Y_{beam} > 0.1$) has been defined as projectile like (PL), whereas, 
($Y_{c.m.}/Y_{beam} < -0.1$) constitutes the target like (TL)
particles.\\
\begin{figure}
\includegraphics{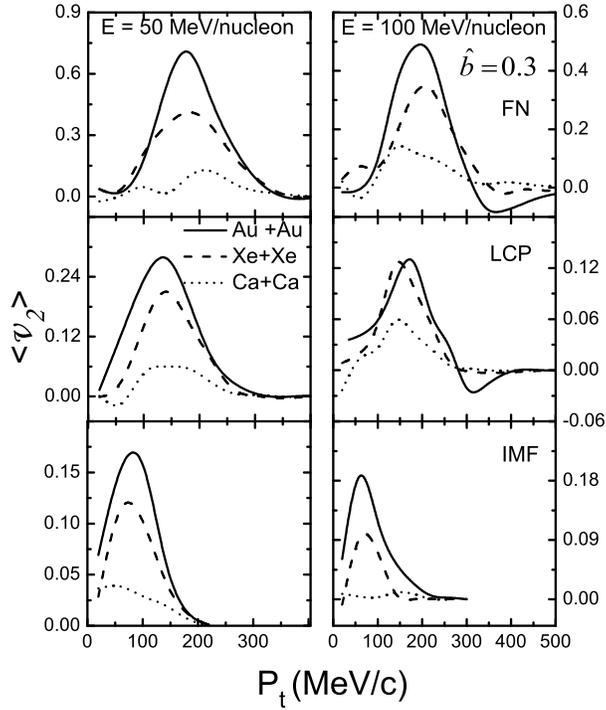}
\caption{\label{fig:1}The transverse momentum dependence of the elliptical flow, 
summed over entire rapidity distribution, at $\hat{b} = 0.3$ for different symmetric reactions at 50 (left) and 100 MeV/nucleon (right) respectively. The top, middle and bottom panels are representing the free nucleons (FN),
light charged particles (LCP's) and intermediate mass fragments (IMF's), respectively.}
\end{figure}
\begin{figure}
\includegraphics{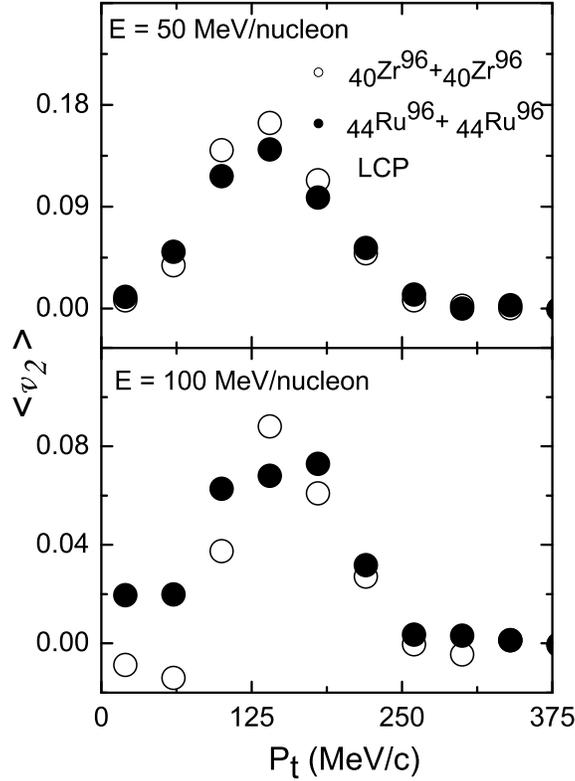}
\caption{\label{fig:2}The transverse momentum dependence of the elliptical flow,
summed over entire rapidity distribution, for 
LCP's at  50 (top) and 100 MeV/nucleon (bottom), respectively. 
The reactions under study having same mass 
number and different 
atomic number. 
The reactions are analyzed with MSTM algorithm.}
\end{figure}
\begin{figure*}
\includegraphics{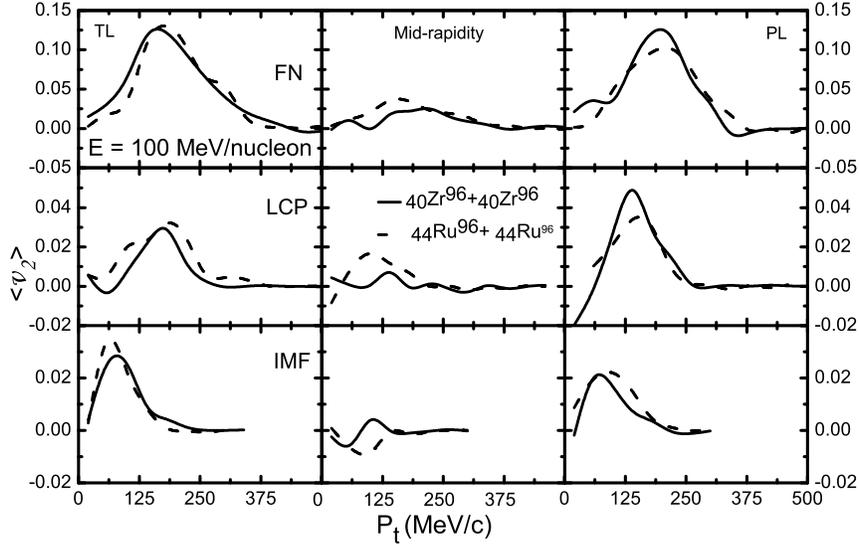}
\caption{\label{fig:3}The transverse momentum dependence of the elliptical flow 
at $E = 100$ MeV/nucleon for the 
reactions displayed in Fig.\ref{fig:2}. The left, middle and right panels are representing 
target-like, mid-rapidity and projectile like distributions, respectively, while, top, middle
and and bottom panels have same meaning as that of Fig. \ref{fig:1}. The reactions are 
analyzed with MST 
algorithm.}
\end{figure*}

\begin{figure}
\includegraphics{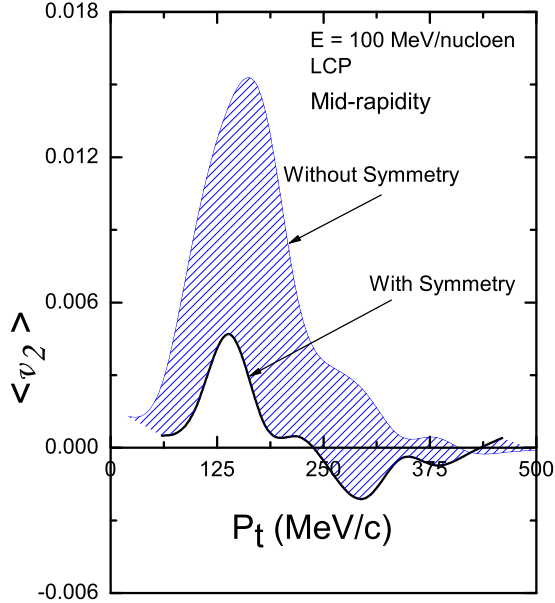}
\caption{\label{fig:4}The transverse momentum dependence of elliptical flow for 
LCP's in the mid-rapidity region at 
E = 100 MeV/nucleon. The panel is exhibiting 
the effect of symmetry energy on the $_{40}Zr^{96}+_{40}Zr^{96}$ reaction.}
\end{figure}

\begin{figure}
\includegraphics{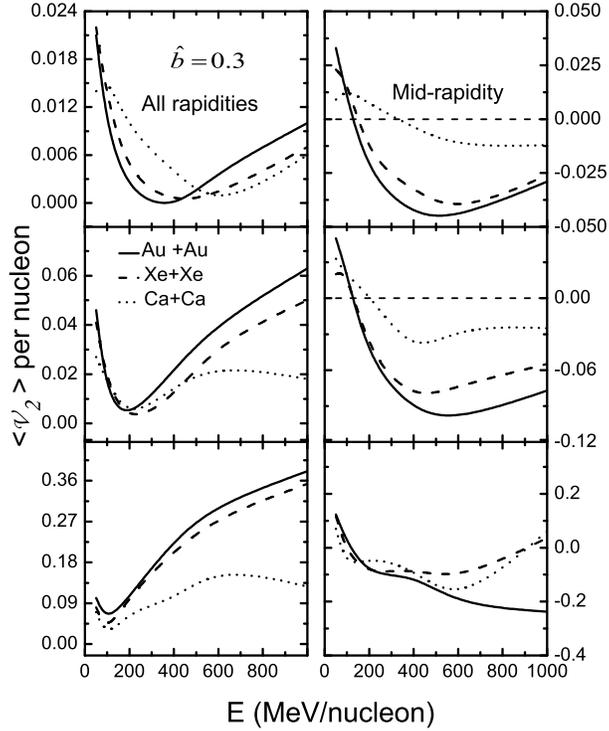}
\caption{\label{fig:5}The variation of the elliptical flow,
summed over entire transverse momentum, with beam energy 
at $\hat{b} = 0.3$ for different symmetric reactions
over entire rapidity range and at mid-rapidity in the left and right panels, respectively.
The top, middle and bottom panels have same meaning as that of Fig.\ref{fig:1}.}
\end{figure}
\begin{figure*}
\includegraphics{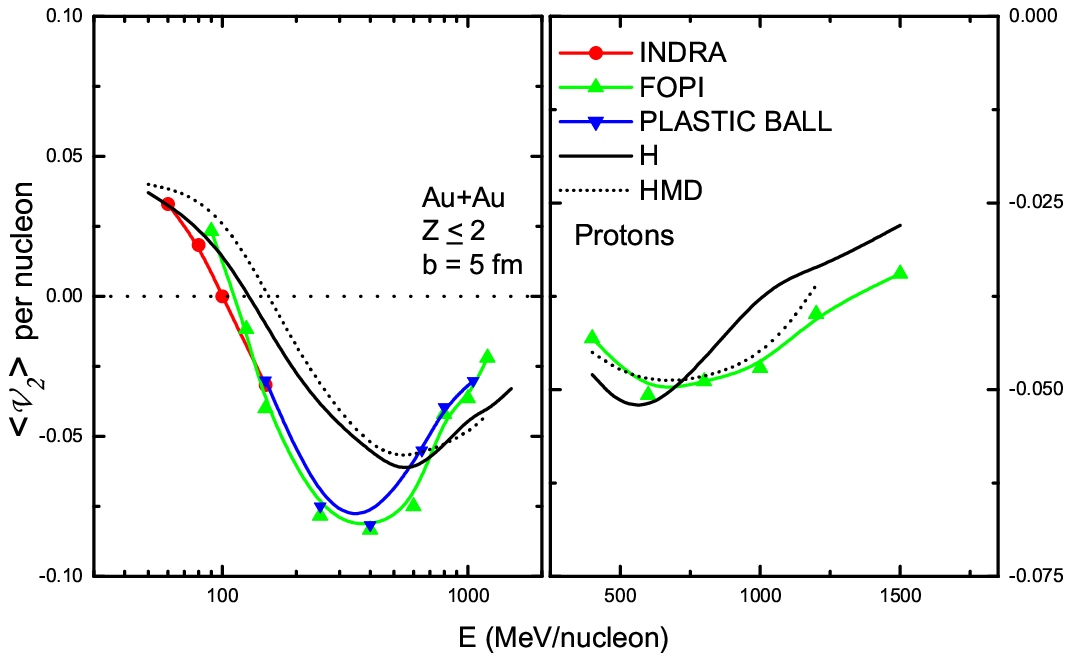}
\caption{\label{fig:6}The variation of the elliptical flow,
summed over entire transverse momentum, with beam energy 
at $|y|$ = $|\frac{y_{c.m}}{y_{beam}}| \le 0.1$ 
for $_{79}Au^{197}+_{79}Au^{197}$
reaction. Here comparison is shown with experimental findings of INDRA, FOPI and PLASTIC BALL 
Collaborations\cite{Luka05,Andr05,Luka04}.}
\end{figure*}
In Fig.\ref{fig:1}, the final state elliptical flow is displayed for the free 
particles (upper panel),
light charged particles (LCP's) [$2\le A \le 4$](middle),
and intermediate mass fragments (IMF's) [$5\le A \le A_{tot}/6$](lower panel) 
as a function of transverse momentum 
($P_t$). A Gaussian type behavior is observed in all cases.  Note that this 
elliptical flow is integrated over entire rapidity range. This
Gaussian type behavior is quite similar to the one obtained by Colona and Toro 
{\it et al.}\cite{Colo07}. 
One sees that elliptical flow is positive 
in the whole range of $P_t$. The collective rotation is one of the main mechanism to 
induce the positive
elliptical flow \cite{Ma93}. It is 
also evident from the figure that the peaks of the Gaussian shifts toward lower 
values of $P_t$ for heavier fragments.
This is due to the fact that the free and light charged particles feel the mean field 
directly, while heavy fragments 
have weaker sensitivity \cite{Yan07}. 
Furthermore, the peak values of $v_2$ for the free 
nucleons and LCP's at 50 MeV/nucleon is 0.70, 0.411, 0.126 and 0.27, 0.20, 0.059 for the 
reactions of $_{79}Au^{197}~+~_{79}Au^{197},~_{54}Xe^{131}~+~_{54}Xe^{131}$ 
and $_{20}Ca^{40}~+~_{20}Ca^{40}$, respectively and the corresponding 
ratios are $\approx$ 5.0, 3.3 and 1. The mass ratio
of these reactions is 4.93, 3.27 and 1, whereas, $N/Z$ ratio is 1.49, 1.42 and 1. 
The $v_2$ ratios are 
in closer agreement with the system mass ratios. The results, however, are different at
E = 100 MeV/nucleon. Note that the peak values for the free nucleon are 
0.48, 0.34, 0.134 and for LCP's numbers, are 0.132, 0.125, 0.058. Their corresponding
ratios are $\approx$ 2.92, 2.36 and 1, indicating a clear deviation
from the mass ratio. \\ 
To further strengthen our interpretation of the estimated $v_2$ ratios, 
we display in Fig.\ref{fig:2}, the reactions of 
$_{40}Zr^{96}~+~_{40}Zr^{96}$ and
$_{44}Ru^{96}~+~_{44}Ru^{96}$ under the same conditions for LCP's. 
These reactions are analyzed within MST method with momentum cut. 
Interestingly, the $N/Z$ effect is 
more visible at
E = 100 MeV/nucleon, indicating that this difference is not due to the mass dependence
alone,
but is due to isospin effect also.
Our findings are also supported by Zhang {\it et al.} \cite{Zhan00}, where they 
showed that neutron-rich
system exhibits weaker squeeze-out flow. At low incident energy (say 
50 MeV/nucleon), binary collisions are rare, therefore isospin in the mean field 
does not play role. On the other hand, around 100 MeV/nucleon, both isospin of the 
mean field and binary collisions contribute, making isospin maximum. At higher incident 
energies, the role of mean field reduces. This situation is similar to the 
intermediate mass fragments, where maximum value is obtained around 100 MeV/nucleon
\cite{Tsan93}.\\
To further 
understand the origin of this isospin effect, the 
transverse momentum dependence of elliptical flow for target-like, 
mid-rapidity  and
projectile-like distributions is displayed in Fig.\ref{fig:3}. From the figure, we see that isospin effect 
originates from the mid-rapidity region or in other words from
the participant zone. It is also clear that the isospin effects 
are stronger for LCP's compared to 
other fragments. This is due to the fact that heavier fragments have weak 
sensitivity towards  
mean field \cite{Yan07}.\\
In Fig.\ref{fig:4},
we display the transverse momentum dependence of elliptical flow 
for LCP's in the mid-rapidity region with and without symmetry energy. 
The effect of symmetry energy is clearly visible in the figure.  
This is in agreement with the 
findings of Chen {\it et al.} \cite{Chen03}, where it was concluded that 
light clusters production acts as a probe for  
symmetry energy. This is strengthening our agreement that elliptical flow 
depends on the $N/Z$ ratio or alternatively the isospin dependence rather than
on the size of the interacting system. \\
\begin{figure}
\includegraphics{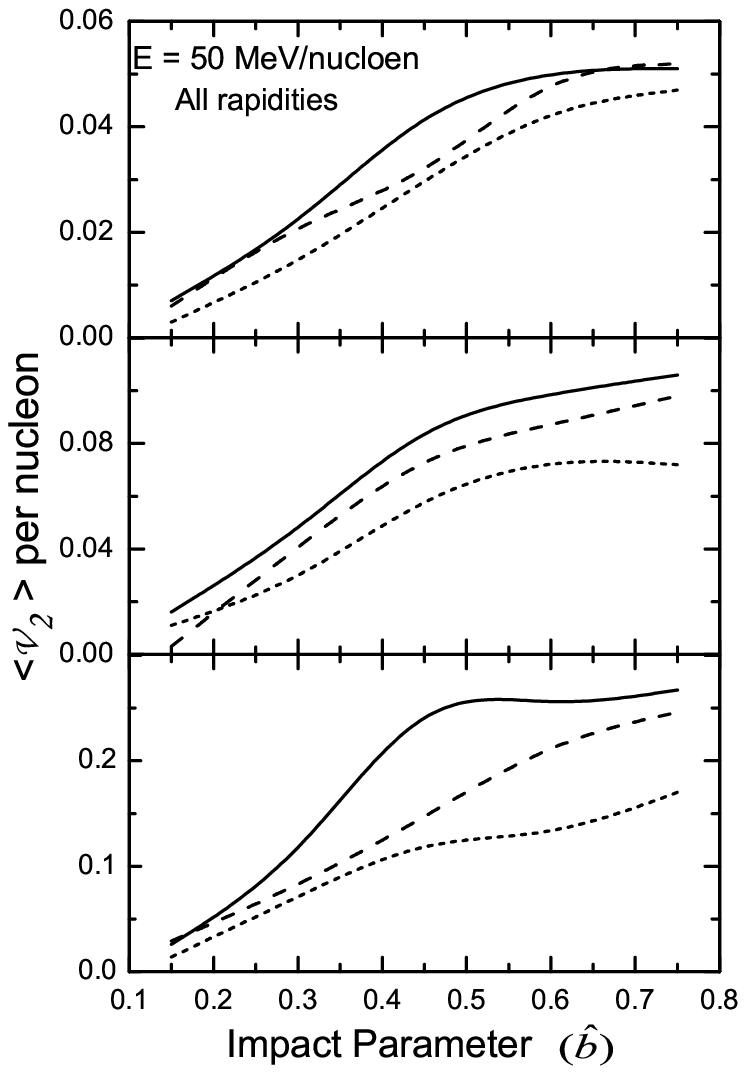}
\caption{\label{fig:7}The impact parameter dependence of the elliptical flow, 
summed over entire transverse momentum and rapidity distribution, at incident
energies 50 MeV/nucleon. The top, middle and bottom
panels are for free particles, LCP's and IMF's, respectively.}
\end{figure}
In Fig.\ref{fig:5}, we display the variation of the excitation function of 
elliptical flow $v_2$ for free, LCP's 
and IMF's over entire rapidity and mid-rapidity region. 
The elliptical flow is found to become less positive 
(entire rapidity) or more negative (mid-rapidity) with the increase in 
the beam energy, upto a certain energy, and then again
becomes more positive or less negative. This is due to the fact that spectators
 move faster after the $v_2$ has reached 
a minimum value\cite{Andr05}.  This energy, at which the behavior
changes is found to 
decrease with the size of the fragment. It
means that the flow of heavier fragments is larger compared to LCP's/free
nucleons at all beam energies. These type of findings are also reported
by different authors in Ref. \cite{Zhan02}. This is true for entire rapidity 
region as well as for mid-rapidity region. \\
The interesting phenomena of transition from in-plane to out-of-plane is 
observed at mid-rapidity region\cite{Luka05, Zhan06}, while no transition 
is observed when integrated over  
entire rapidity region. The energy at which this transition is observed is 
dubbed as transition energy($E_{Trans}$). 
It means that participant zone is responsible for the transition from in-plane to 
out-of-plane. That's why free particles and LCP's, which originate from 
the participant zone, are showing a 
systematic behavior with the beam energy as well as with the composite mass of the system. 
The elliptical flow for these particles is found to become more 
negative with the increase in the composite mass of the system. 
Heavier is the system, more is the 
Coulomb repulsion and more negative is the elliptical flow.
This systematics of $E_{Trans}$ with composite mass of the system is discussed later.\\  
In Fig. \ref{fig:6}, we show $v_2$@mid rapidity ($|y|$ = $|\frac{y_{c.m}}{y_{beam}}| \le 0.1$) 
for $Z \le 2$ (left panel) and for protons (right panel) as a function of the incident energy. 
The rapidity cut is in accordance with the experimental findings.
The theoretical results are compared with 
the experimental data extracted by INDRA, FOPI and PLASTIC BALL 
collaborations\cite{Luka05,Andr05,Luka04}.
With the increase in the incident energy, elliptical
flow $v_2$ changes from positive to negative values exhibiting a transition from 
the in-plane to 
out-of-plane emission of nucleons.  This is because of the fact that 
the mean field, which contributes to the formation 
of a rotating compound system, becomes less important and the collective 
expansion process based on the 
nucleon-nucleon scattering starts to be predominant. This competition between the 
mean field and N-N collisions depends strongly on the effective interactions that leads to 
the different transition energy due to different equations of state. 
Due to repulsive nature of 
the momentum dependent interactions, which leads to the suppression of 
binary collisions, less squeeze-out is observed in the presence of 
momentum dependent interactions
(HMD) compared to static one (H). 
The maximal negative value of $v_2$ is obtained around $E = 500$ 
MeV/nucleon with hard (H) and hard momentum dependent (HMD) equations of state.
This out-of-plane emission decreases again towards the higher incident energies. 
This happens due to faster movement of the spectator matter after 
$v_2$ reaches the maximal negative value \cite{Andr05}.
This trend is in agreement with experimental findings. A close agreement with data 
is obtained in the presence of hard equation of state for $Z \le 2$ particles, while, 
in the presence of momentum 
dependent interactions for protons. 
Similar results and trends  have also been reported by Zhang et.al. in their
recent communication \cite{Zhan06}.\\
\begin{figure}
\includegraphics{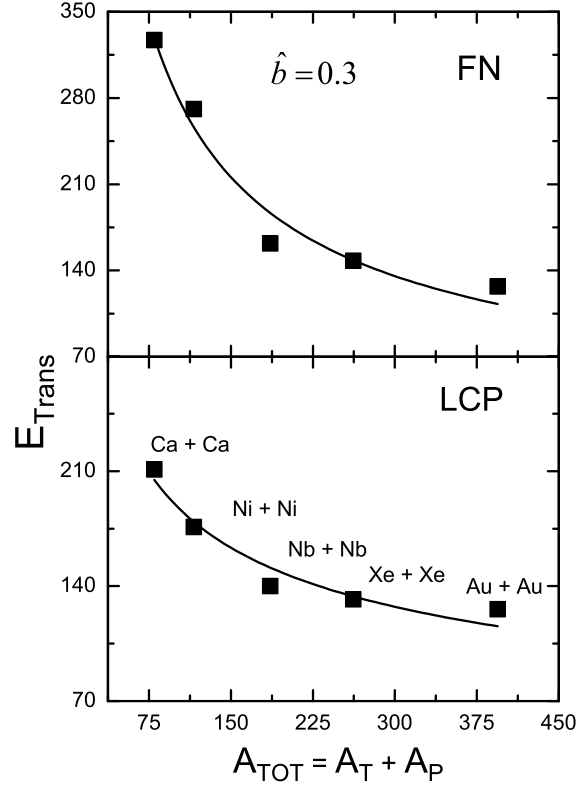}
\caption{\label{fig:8}The transition energies for elliptical flow at intermediate 
energies as a function of the combined mass
of the system. The upper panel is for the free nucleons, while, lower panel is for the LCP's.}
\end{figure}
The investigation of the elliptical flow with scaled impact parameter over 
entire rapidity range is 
displayed in Fig.\ref{fig:7}.
Here the top, middle and bottom panels represent the free nucleons, LCP's and
IMF's. 
The value of the elliptical flow $v_2$ becomes more positive with the impact parameter 
and composite mass of the system 
at $E = 50 $ MeV/nucleon, while at higher energies (not shown here), 
it is found to become less positive (entire rapidity) or more negative (mid-rapidity) with composite mass of the system. 
This is indicating the dominance of 
the in-plane
flow at low incident energies with increasing impact parameter and 
composite mass of the system. Moreover, dominance of 
the out-of-plane flow at higher energies with small impact parameter and 
composite mass of the system is observed.  
With the increase in the beam energy, the expansion of the compressed zone 
becomes more vigorous, while, 
with an increase in the impact parameter, participant zone decreases, resulting an  
increases in the spectator region indicating 
dominance of azimuthal anisotropy with impact parameter. On the other hand, it reduces with beam energy. 
These observations are 
consistent with the experimental
findings and with other theoretical works \cite{Zhan02,Ma93,Pete90}.\\
Finally, we carry out the system size dependence of the elliptical flow for 
free nucleons and LCP's. In Fig.\ref{fig:8},
we show the transition energy $E_{Trans}$ as a function of the composite mass of the 
system for free nucleons and LCP's.
From the figure, we see that the transition energy decreases with the 
composite mass of the system as well as with the size of the fragment.
The reason for this is that the pressure produced by the 
Coloumb interactions increases with the system size.
This dependence can be fitted using a power law of the kind:   
\begin{equation}
E_{Trans} = C(A_{tot}^{-\tau})
\end{equation}
The exponent $\tau$ is found to be 2 times for free particles (0.67) 
compared to LCP's (0.35). This
exponent is quite smaller compared the exponent of balance energy in 
directed flow \cite{Zhan06}. This is due to the different origin of the balance and 
transition energy. 
The balance energy counter balances the mean field and N-N collisions, 
while transition energy is due to the more 
complex effects such as expansion of the compressed zone and shadowing of 
the cold spectator matter.\\
\section{Conclusion} 
In conclusion, we have investigated the elliptical flow of fragments for different 
reacting systems at incident energies between 50 and 1000 MeV/nucleon
using isospin-dependent quantum molecular dynamics (IQMD) model. The elliptical flow is found to show a transition 
from in-plane to out-of-plane at a certain beam energy in mid-rapidity region, while no such transition is 
observed when integrated over entire rapidity region. This transition energy is 
found to decrease with the composite
mass as well as with the size of the fragment. The transition energy is further 
parametrized in term of mass power law. In addition, LCP's exhibit 
isospin effect in the mid-rapidity region.\\

\begin{acknowledgments}
This work has been supported by the Grant no. 03(1062)06/ EMR-II, from the Council of Scientific and
Industrial Research (CSIR) New Delhi, Govt. of India.\\
\end{acknowledgments}

\end{document}